\documentclass[runningheads, a4paper]{llncs}
\usepackage{amssymb}
\usepackage{graphicx}
\usepackage{url}
\usepackage{algorithmic}
\usepackage{algorithm}
\usepackage{float}
\floatname{algorithm}{Program}

\usepackage{booktabs}
\usepackage{arydshln}

\newcommand{\SWITCH}[1]{\STATE \textbf{switch} (#1)}
\newcommand{\ENDSWITCH}{\STATE \textbf{end switch}}
\newcommand{\CASE}[1]{\STATE \textbf{case} #1\textbf{:} \begin{ALC@g}}
\newcommand{\ENDCASE}{\end{ALC@g}}

\newcommand{\DEFAULT}{\STATE \textbf{default:} \begin{ALC@g}}
\newcommand{\ENDDEFAULT}{\end{ALC@g}}
\newcommand{\DEFAULTLINE}[1]{\STATE \textbf{default:} }

\title{Programmable Restoration Granularity in Constraint Programming}

\begin{document}

\author{Yong Lin and Martin Henz}

\titlerunning{}

\institute{
School of Computing, National University of Singapore, Singapore\\
\email{\{linyong, henz\}@comp.nus.edu.sg}
}

\date{}
\maketitle

\begin{abstract}
    
In most constraint programming systems, a limited number of search engines is offered while the programming of user-customized search algorithms requires low-level efforts, which complicates the deployment of such algorithms. To alleviate this limitation, concepts such as computation spaces have been developed. Computation spaces provide a coarse-grained restoration mechanism, because they store all information contained in a search tree node. Other granularities are possible, and in this paper we make the case for dynamically adapting the restoration granularity during search. In order to elucidate programmable restoration granularity, we present restoration as an aspect of a constraint programming system, using the model of aspect-oriented programming. A proof-of-concept implementation using Gecode shows promising results. 
 
\end{abstract}

\section{Introduction}

Constraint Programming (CP) solves combinatorial problems through constraint-based search, which explores a problem-defined search tree in a certain order. The exploring of search tree may encounter a failure, which signals a false search direction. To recover search from such a failure, restoration service is called to provide an ever visited state to switch search direction.
 
 
 In CP systems, restoration has been implemented in several manners. Mozart/Oz system \emph{copies} each ever accessed state for direct retrieval, in a format of \emph{computation space} (space for short); Constraint Logic Programming (CLP) systems such as CHIP~\cite{CHIP}, clp(FD)~\cite{clpfd} and ECL$^i$PS$^e$~\cite{eclipse} record state change information in a \emph{trail} data strucutre to roll back; Gecode~\cite{Gecode} system maintains generated \emph{constraints} to recompute~\cite{constraintService}; our newly proposed recollection~\cite{prototype} stores constraint propagation affected variable \emph{domains} to recollect. 
 
 For all above implemented techniques, they differ on the granularity of the stored information. Copying is coarse-grained since a computation space encapsulates all speculative computation  structures; on the other hand, recomputation does not store any specific data structures, but the instructions committed for guiding search. Both trailing and recollection are finer-grained than space copying since they store only part information of a state. Actually, the granularity of a restored technique determines its advantages and limitations. If one can program to synthetically use these techniques in a system, more efficient restoration is expected; an early effort was conducted by the proposal of computation space, which enables user to place state copies in the search tree as they intend. However, as we have claimed, space is coarse-grained, while recomputation is only aware of committed constraints. Therefore, we advocate that it is a non-trivial study to include other finer-grained techniques to program a more powerful restoration scheme.
 

 In this paper, we propose to program restoration granularities, aiming at improving constraint programming system performance. The initial effort has been conducted by integrating coarse-grained copying, finer-grained recollection and constraint-aware recomputation. Especially, we propose to program restoration granularity in aspect-oriented programming for a more modular and flexible restoration implementation. The rest of this paper is organized as follows: Section~\ref{basics} briefly goes through the notions in CP systen implementation; in Section~\ref{sample}, we present a prototype of programming restoration granularity ; subsequently, in Section~\ref{program}, we employ the concept \emph{aspect} to illustrate the landscape of more modular and flexible restoration implementation, taking Gecode as an example; last, we conclude in Section~\ref{conclusion}.

\section{Basics} \label{basics}

In CP systems, constraint propagation is demonstrated insufficient to solve a problem when it reaches a fix point. Search then comes into play; it branches on the fix point to generate constraints and then commits one of generated constraints to proceed. This interleaving of constraint propagation and branching creates a search tree, where each node is a computation state. In such a search tree, branches are constraints, internal nodes are fix points and leaf nodes are either solutions or search failures. A search failure indicates a false search direction, which requires the restoration of a previously visited internal state to switch search directions. 

In constraint programming systems, state restoration can be implemented in several ways. \emph{Trailing}-based method stores how one states was modified to reach other conjunct state in a \emph{trail} structure to \emph{roll back} previous performed operations~\cite{clpSurvey}; this method is efficient for the problems imposing weak propagation~\cite{comparing}. However, trailing couples tightly with propagation engine and search facilities, which potentially limits parallel exploration. \emph{Copying} approach duplicates each ever accessed fix point state and restoration is a simple direct state retrieval. This technique causes neither runtime nor memory issues for small and medium size problem. As for large problems, copying may introduce memory management penalty on account of its substantially occupied memory~\cite{comparing}. Recomputation barely maintains previously committed constraints so that it can \emph{recompute} a state. This approach consumes rather limited memory, but its runtime may be dragged significantly if a problem is computation expensive. Our proposed \emph{recollection} memorizes the variables that were changed in fix point reasoning steps at a moderate extra memory investment; it conducts restoration by updating a higher level state, using the stored variables. For all these restoration techniques, we can observe that their ways to conduct restoration are determined by the granularity of information they have stored during search. Specifically, copying is coarse-grained since it stores all information of states; by contrast, trailing and recollection are finer-grained. As for recomputation, it merely stores the meta-information (constraints) that was used to instruct search directions.  

\section{A Prototype} \label{sample}

In this section, we would present a proof-of-concept that programs restoration granularity. The organization of this section is as follows: Section~\ref{motivation} highlights the traits of a few problem, which spurs our investigation in programming restoration granularity and we describe the prototype in Section~\ref{ideas}; evaluation is conducted in Section~\ref{evaluation}.

  \subsection{Motivation}~\label{motivation}
  
   A deep search tree may be expanded for solving a hard problem. Table~\ref{stat} illustrates the search tree statistics of four problems exploring for the first solution, where the Queens problem is modeled by either a set of disequality constraints or three global constraints of the ``all-different constraints" family (denoted as Queens-S). In this table, the column \emph{failures} counts the total number of failures during search; \emph{first} signals the tree level where the first search failure emerges, while \emph{peak} is the value of the deepest tree level. [1, \emph{first}) accumulates the number of failures occurs between the root and the \emph{first} level, and [\emph{first}, \emph{peak}] records the number of failures between \emph{first} and \emph{peak}.

  \begin{table}[h!]
\begin{center}
\begin{tabular}{l  c c c  c c c c c c c }
  \toprule
  Problem & & \emph{failures} && \emph{first} && \emph{peak} && [1, \emph{first}) && [\emph{first, peak}]\\
  \midrule
 Queens(200) && 146,838 && 164&& 200 &&  0 && 146,838 \\
 Queens-S(200) && 146,838 && 164 && 200 && 0 && 146,838 \\
 Knights(22) && 19,877 && 386 && 451&& 0 && 19,877 \\ 
 Sport-League(22) && 1,035 &&62 && 249 && 5 && 1,030 \\
  \bottomrule 
\end{tabular}
\end{center}
\caption{Search Tree Statistics of Problem Search Trees}
\label{stat}
\end{table}%

Although any of restoration techniques is capable of accomplishing its duty correctly, they respectively exhibit advantage towards certain case, as we have claimed. From these search tree statistics, we perceive that the emergence of the first failure can be an important signal for more intensive search failures. Therefore, if copying is employed as an underlying restoration to support the solving of these problems, then the space copies maintained between the root  $first-1$ level cannot contribute at all. This indicates that an ideal restoration should customize problem search traits. 
 
\subsection{Programming Granularity}    \label{ideas}
 
 To deal with the skewed failure distribution as displayed in Table~\ref{stat}, a straightforward method is storing restoration information of difference granularity as search proceeds, and restoration then adapt the stored information to conduct relevant process. We initially program a prototype by integrating copying, recomputation and recollection. In this prototype, we take the search tree level where the \emph{first} failure emerges as a \emph{border level}, which horizontally divides the search tree into \emph{upper} and \emph{bottom} two parts. For search in the upper part of the search tree, only generated constraints are kept, while neither coarse-grained state copies nor finer-grained propagation modified variables is stored. When search explores beyond the border level, both copying and recollection will be activated to collaborate: the placing of a coarse-grained state copy alternates \emph{n} finer-grained recollection explorations (\emph{n} is configurable and set to \emph{eithgt}). Because of this change of information granularity, restoration now need to switch between different code segment. Specifically, when a state in upper part search tree is expected, the restoration routine switches to recomputation. When a state in the bottom part is required, the restoration routine first attempts to recollect from the nearest state copy; if this effort fails, it then recomputes to restore the border state first and then update this border state to the requested one by recollection. We can take an optimization measure by maintaining a border state copy to avoid the possible recomputation when restoring a bottom state. Similarly, one can also employ other measures to optimize the recomputation in upper part.
 
  
\subsection{Evaluation}    \label{evaluation}
  
 We evaluate our programmed prototype over the four problems, where we sought the motivation to program restoration granularity. Both recomputation and recollection enable adaptive function and set \emph{copying distance} to \emph{eight}. The evaluations were conducted on a Intel Core 2 Quad processor PC system, running an Ubuntu operating system 11.10 with four Gigabyte main memory. We built our sample on top of Gecode version 3.7.3, which also served as the reference instance of comparisons. Each collected runtime\footnote{We take wall clock time.} value is an arithmetic mean of 20 runs with a variation coefficient less than 2\%; memory numbers are peak memory consumption.
 
  \begin{table}[h!]
\begin{center}
\begin{tabular}{l c c c c c c c c }
\toprule
 & \multicolumn{2}{c}{Recomputation} & & \multicolumn{2}{c}{Recollection} & & \multicolumn{2}{c}{ Prototype} \\
\cline{2-3} \cline{5-6 }\cline{8-9}
   Problems & Time(ms) & Mem(KB) & & Time(ms) & Mem(KB) & & Time(ms) & Mem(KB)  \\
\midrule
 Queens(200) & 4,330 & 25,748 &  & 4,578 & 28,238  &  & 4,601& 6,244 \\
 Queens-S(200) &  2,156 & 1,485 &  & 2,473 & 3,974 &   & 2,469 & 542 \\
 Knights(22) & 1,858 & 4,460 &  & 1,704 & 4,592  &  & 1,744 & 2,333 \\ 
 Sport-League(22) & 352 & 7,710& & 331 & 7,937 & & 339 & 6,109 \\

\bottomrule
\end{tabular}
\end{center}
 \caption{Comparisons of Prototype with Recomputation and Recollection}
 \label{results}
\end{table}

We compare the prototype with both adaptive \emph{recomputation} and adaptive \emph{recollection}, and Table~\ref{results} depicts the evaluations results. These numbers reveal that our prototype can significantly save memory than the other two restoration alternatives; but for Sport-League problem, the memory saving is not as significant, which can be on account of it \emph{first} failure comes earlier (at level 62 of a tree with peak depth 249). Meanwhile, Queens problems expose better runtime performances by adaptive recomputation. Nevertheless, it deserves to highlight that Knights problem almost halves memory consumption than the other two techniques, while marginally improves its runtime than recomputation. This promising result confirms the opportunities of programming restoration granularity to seek better performance.

\section{Programmability} \label{program}

 As mentioned in previous section, the way to implement restoration is actually determined by the granularity of storing information for restoration. This information is implemented to store by search engine at exploration steps; on the other hand, restoration is a service provided for search engine. This observation claims that the maintained restoration information \emph{cuts across} the two abstractions: search engine and restoration. 
   
  In developing applications, the occurrence of \emph{crosscutting} abstraction is not rare; transactions, security-related operation, logging etc all exemplify crosscutting abstractions. To facilitate the programming of crosscutting abstraction, aspect-oriented programming~\cite{aspect}(AOP) proposes to encapsulate a crosscutting abstraction as an \emph{aspect}.  The implementation of an aspect mainly consists of two tasks: \emph{advice} and \emph{pointcut}. An advice is a means of specifying the code to run at a \emph{joint point}, where additional behaviors attempt to add in the program; a pointcut determines the matches of executing specified advice at a joint point.    
  
  To program restoration granularity, one should correctly implement the switching between restoration technique code segments. In our prototype, we achieve the communication between code segments by introducing a signal. However, this signal couples tightly with a specific program; suppose one redefines the criteria to switch restoration techniques, the code is quite likely to change, probably drastically. Therefore, it is of great significance to implement a more flexible and modular design to program restoration granularity, and we claim that programming restoration as an aspect can achieve this design goal. 
  
  In this section, we propose to adopt aspect-oriented programming to program restoration granularity in Gecode system for a modular and flexible implementation. Specifically, Section~\ref{facilities} briefly explains the design of search facilities in Gecode system; In Section~\ref{programming}, we discuss the issues related with deploying aspect in Gecode search facilities.
  
 \subsection{Search Facilities}  \label{facilities}
  
  In the Gecode system, it defines a class of \emph{Edge} to keep the restoration information of each visited fixpoint; all Edge objects will be pushed into a stack structure \emph{Path} by search engine.  Program~\ref{edge} outlines the class declaration of Edge. In the Edge, the \emph{Choice} object encapsulates fix point generated two constraints\footnote{We restrict our discussion on binary search tree}, which are respectively represented by 0 and 1; the currently committed constraint alternative is recorded by the integer variable \emph{\_alternative}. Choice object is a compulsory information component in each created Edge object since it instructs search direction by committing constraint, whereas the space copy \emph{\_space} is optional. Specifically, if search engine wraps a space in each Edge, then it is a copying-based restoration; if none of Edge stores space copy, it is a recomputation-based restoration; hybrid variants can be obtained by placing space copies in Edges occasionally such as adaptive recomputation. In Gecode, all restoration related services are defined on the structure \emph{path}, which collects the Edge objects generated between root and current node.    

 \begin{algorithm} [h!]
 \scriptsize
  \begin{algorithmic}[100S] 
      \STATE \textbf{class} Edge  \hspace{3mm}   \{
      \STATE   \hspace*{6mm} Space * \_\emph{space};  \hspace*{3mm}  /* Space copy */
      \STATE    \hspace*{6mm}  Choice * \_\emph{choice};  \hspace*{3mm}/* fixpoint generated Choice */
      \STATE    \hspace*{6mm} \textbf{unsigned int} \_\emph{altternative};\hspace*{3mm} /* committed choice alternative */
       \STATE  \hspace*{6mm} vector<\textbf{int}> \_\emph{doms};  \hspace*{3mm}/* variable domains */
      \STATE  	\hspace*{3mm}\textbf{public}:
      \STATE  \hspace*{6mm}Edge(Space * s, Choice * c, vector<int> \emph{\_doms}):  \hspace*{3mm}\{
      \STATE   \hspace*{6mm} \_\emph{alternative} $= 0$;  \hspace*{3mm} /* commit to the first alternative */
      \STATE   \hspace*{6mm} $\dots$  \hspace*{3mm} /* other initialization statements */
      \STATE \hspace*{6mm}\}
      \STATE \hspace*{6mm}$\dots$
      \STATE \hspace*{3mm}\}
    \end{algorithmic}
     \caption{Class Edge}
      \label{edge}
  \end{algorithm} 
  
As we have claimed, space is coarse-grained data, while constraints are meta-information for instructing search. We introduce a finer-grained data structure \emph{\_dom}, which collects the variable domains that were affected in the fix point reasoning\footnote{We insist on Finite Domain Integer Domain problems in this discussion}. This vector contains sufficient information to conduct recollection. One can also introduce other data structures to further explore other restoration possibilities.

  In Gecode, search engine is constructed by programming on computation spaces. The space provides an interface status() to trigger internal constraint propagation, which returns a status value; search engine then switches to the relevant code segment according to the space status, as depicted in Program\ref{engine}.  If the space return a constraint propagation results of failure, the search engine first requests to adjust search direction by calling Adjust()\footnote{It will return a Boolean false if another search direction is impossible}, and then enter the code segment where restoration is defined\footnote{Restore() method is not actually defined in search engine, we expose it body in search engine for facilitating explanation}. 
  
  \begin{algorithm} [h!]
 \scriptsize
  \begin{algorithmic}[100S] 
    \WHILE{\emph{\textbf{true}}}
      \SWITCH {Status(\emph{space})}  \hspace{3mm} /* query space status */
      \CASE { \textbf{\emph{fixpoint}}  }
       \STATE Choice \emph{ch}$\leftarrow$Choice(\emph{space})  \hspace{3mm}   /* return solution space */   
       \STATE Push(Edge(\emph{ch} $\dots$))
       \STATE Commit(\emph{space}, \emph{ch}) \hspace{3mm} /* commit a constraint to space */
        \ENDCASE      
      \CASE  {\textbf{\emph{solution}}}
        \RETURN \emph{space}  \hspace{3mm} /* return solution space */  
        \ENDCASE
        \CASE {\textbf{\emph{failure}}}
        \IF { \emph{\textbf{not}} Adjust($\dots$)}       
        \STATE   \textbf{break}   \hspace{3mm}  /* The problem is not solvable */    
        \ENDIF   
        \STATE Restore($\dots$)  \{
        \STATE \hspace{1.5mm}\emph{Pointcut1}   
        \STATE \hspace{1.5mm}Recomputation code   \hspace{3mm} /* originally programmed recomputation */
        \STATE \hspace{1.5mm}\emph{Pointcut2} \}      
        \ENDCASE
      \ENDSWITCH
      \ENDWHILE
    \end{algorithmic}
     \caption{Search Engine}
      \label{engine}
  \end{algorithm} 
  
 \subsection{Restoration as an Aspect} \label{programming}  
  
  To program in aspect-oriented programming paradigm, it requires support from the underlying programming language compiler/system. Currently, quite a few programming languages have implemented AOP, within the language, or as an external library; AspectC++~\cite{aspectcpp} has provided a set of extensions to facilitate the use of AOP in C++. Gecode is an open source constraint programming library constructed in C++; thus, a following exploration is provisioning restoration service of Gecode with the aspect-oriented programming paradigm. 
  
  To deploy restoration as an aspect, it requires to address the definition of advice and pointcut. Specifically, we define advice to exploit the information of a specific granularity so that it can individually, or collaboratively with the recomputation code, restore ever visited fix point; this collaboration requires the pointcuts (advice can be matched and applied for execution) either before and after the orignal recomputation code segment, as illustrated in Program~\ref{engine}. In our prototype, we place the code of copying and recollection at the place marked by \emph{pointcut2}; but as we have stressed, we currently use a specific signal to coordinate the code switching. Thus, we conjecture that a more modular and flexible restoration can be built by using AOP. 
  
   \section{Conclusion} \label{conclusion}
   
   Restoration is a key component of search facilities in constraint programming systems; its implementation can significantly impact on the performance and construction of a search engine. Although the proposal of computation space facilitates the programming of restoration, it is coarse-grained since a space encapsulates all data structures of a state in the search tree. 
   
  In this paper, we have proposed to program restoration granularity. We implemented such a prototype by integrates coarse-grained copying, finer-grained recollection and constraint-based recomputation; its evaluation gives promising results. In the prototype, we employed a specific signal to coordinate the execution of between restoration code segments, which couples tightly with a program. Thus, we have further propose to provision restoration with aspect-oriented programming. The significance of this prospective effort is, on the one hand, modularizing restoration code segments that implement different techniques while flexing their assemble. On the other hand, it also potentially provides more provisions to construct search engines that run a wide spectrum of search algorithms.  
  

 \bibliographystyle{plain}
 \bibliography{references}

\end{document}